\documentclass[twoside]{ilcws10}
\usepackage[latin1]{inputenc}
\usepackage[dvips]{graphicx,epsfig,color}
\usepackage{wrapfig,rotating}
\usepackage{amssymb,amsmath,array}
\usepackage{subfigure}

\begin{document}
\title{Status of the Micromegas semi-DHCAL}

\author{C. Adloff, J. Blaha, J.-J. Blaising, M. Chefdeville, C. Drancourt, A. Espargili\`ere, \\
R. Gaglione, N. Geffroy, Y. Karyotakis, J. Prast and G. Vouters
\thanks{This work was performed within the CALICE collaboration.}
\vspace{.3cm}\\
Laboratoire d'Annecy-le-Vieux de Physique des Particules\\
Universit\'e de Savoie, Annecy-le-Vieux, F-74940 France}

\maketitle

\begin{abstract}

The activities towards the fabrication and test of a 1\,m$^{3}$ semi-digital hadronic calorimeter are reviewed.
The prototype sampling planes would consist of 1\,m$^{2}$ Micromegas chambers with 1\,cm$^{2}$ granularity and embedded 2 bits readout suitable for PFA calorimetry at an ILC detector.\\
The design of the 1\,m$^{2}$ chamber is presented first, followed by an overview of the basic performance of small prototypes.
The basic units composing the 1\,m$^{2}$ chamber are 32$\times$48\,cm$^{2}$ boards with integrated electronics and a micro-mesh.
Results of characterization tests of such boards are shown.
Micromegas as a proportional detector is well suited for semi-digital hadronic calorimetry.
In order to quantify the gain in performance when using one or more thresholds, simulation studies are being carried out, some of which will be reported in this contribution.
\end{abstract}


\section{Introduction}

The goal of the project is the fabrication of a semi-digital hadronic calorimeter (2 bits readout) of 1 m$^{3}$ that can be used to perform measurements of high energy particle shower properties.
At the same time this prototype should be as close as possible to an ILC-like module, fulfilling the constraints on power consumption, geometry and electronic integration.
The calorimeter will be composed of 40 layers of 1.89 cm thick stainless steel absorbers separated by 8 mm planar Micromegas chambers of 1 m$^{2}$ size.
It should be optimized for Particle Flow calorimetry and has therefore a very fine transverse granularity of 1\,cm$^{2}$ \cite{ILC,PFA}.

\section{Description of the 1 m$^{2}$ Micromegas prototype}

\subsection{Mechanical layout}

The 1\,m$^{2}$ Micromegas prototype consists of 6 printed circuit boards of 32$\times$48\,cm$^{2}$ placed in the same gas chamber.
The reason for this arrangement is to avoid a too large energy to be stored in the mesh which could be released in the front-end electronics circuitry during a spark.
The 3\,mm drift gap between the cathode plane and the meshes is defined by means of insulating spacers.
The latter are inserted in the 0.5\,mm wide gap between boards which creates a dead area of 1.6\,\% only.

Copper pads of 1$\times$1\,cm$^{2}$ are first patterned on one side of each board.
In a second time, front-end chips and spark protection circuits are connected on the opposite side.
Finally, an electron multiplying mesh (Bulk Micromegas) is laminated on the pad side \cite{BULK}.
These boards equipped with a micro-mesh on the pad side and front-end chips on the opposite side are called Active Sensor Units or ASUs.
In this scheme, a very small detector thickness of 6\,mm is achieved including 1\,mm of PCB material, 2\,mm for the chips and passive components and 3\,mm of gas. The chamber stainless steel base plate and cover are not taken into account because they will be part of the absorbing medium.

\subsection{Front-end and DAQ electronics}

The anode plane is segmented into 1536 pads of 1$\times$1\,cm$^{2}$ which are wired through the board to the input pins of the front-end chips.
Every chip has 64 channels and is used to read out the signals from an 8$\times$8\,cm$^{2}$ area.
An ASU is equipped with 24 chips intended for processing the detector signals into digital signals.
The latter are coded to 2 bits (3 thresholds) and then routed to one board edge to be read out by a detector interface board (DIF).
An additional board, placed between the ASU and the DIF distributes the detector HV.

The 32$\times$48\,cm$^{2}$ ASUs fabricated so far are equipped with HARDROC2 chips.
These chips are composed of an analog part with a current preamplifier, a fast and a slow shaper and a digital part with 3 discriminators with variable charge thresholds and a 128 event depth memory.
In view of an application at an ILC detector with a 30 millions channels DHCAL, the HARDROC2 can operate in a power-pulsing mode.
The shaping time of the fast shaper being very short \textit{w.r.t.} the detector signal duration, a new chip optimized for the detection of Micromegas signals called DIRAC has been developed \cite{DIRAC}.

\subsection{Operating conditions}

The detector is operated in a mixture of Ar/\textit{i}C$_{4}$H$_{10}$ 95/5.
Ar was chosen for its good ionisation yield (90\,e$^{-}$/cm for MIPs) and low price.
Isobutane is a very efficient UV quencher and thanks to its low ionisation potential, high gains can be achieved at relatively low mesh voltages.
It should be noted that any other standard mixtures such as Ar/CO$_{2}$ or Ar/CH$_{4}$ can be used \cite{AMB}.
In typical working conditions the voltage difference across the 128\,$\mu$m amplification gap is 420\,V.
At such voltage, in Ar/\textit{i}C$_{4}$H$_{10}$ 95/5, a gas gain of about 15000 was measured.
The cathode voltage is set 50\,V higher to provide a large ratio between the amplification and drift field and hence a good electron transparency of the mesh.

\section{Performance of small prototypes}
\label{gassiplex}

The first Micromegas prototypes were three 6$\times$16\,cm$^{2}$ chambers and one of 12$\times$32\,cm$^{2}$ with 3\,mm gas gap and 1$\times$1\,cm$^{2}$ pads. The front-end electronic boards were equipped with Gassiplex chips which perform a measurement of the charge induced on the pads. In contrast with the ASUs, these chambers had the Gassiplex boards connected on the chamber sides, outside the sensitive area. A test structure containing the 4 chambers was tested in 2008 in a 200 GeV/c muon beam at the CERN/SPS facility.

After a selection of events detailed in \cite{AMB}, the Landau charge distribution was measured on all pads showing an MPV of 22\,fC with about 11\,\% variations over all chamber pads. At a charge threshold of 1.5\,fC, an efficiency between 91 and 97\,\% was found for all chambers.
For a given chamber, the efficiency variation was as small as 1\,\% while the hit multiplicity stood between 1.03 and 1.12 depending on the threshold.
At an ILC detector, the capability to separate individual showers within jets will be essential to reach the desired jet energy resolution.
In that respect a low multiplicity as the one measured will help as the detector will add very little to the overlap of shower patterns in the calorimeter.

In 2009, the behaviour of the chambers in 2\,GeV/c electron and hadron showers was studied in the CERN/PS beam lines \cite{PROF}.
The energy deposited in the largest chamber by electron shower secondaries was measured for different number of lead plates placed in front of the stack.
The longitudinal profile can be found in \cite{CALOR} and agrees well with the trend of a Geant4 simulation.

\section{Test of Active Sensor Units for a m$^{2}$ prototype}

ASUs of 32$\times$48\,cm$^{2}$ are the basic units that compose the Micromegas 1\,m$^{2}$ prototype.
In order to first validate the manufacturing process, the signal routing and the overall detector functioning, however, ASU of smaller sizes were first developed. The first ASU had an active area of 8$\times$8\,cm$^{2}$.
It was equipped with a DIRAC chip and irradiated with muons and pions at the CERN facilities in 2008 \cite{RENO}.
Later, 8$\times$32\,cm$^{2}$ chambers with 4 HARDROC1 chips were produced.
Two detector stacks with DIRAC and HARDROC chambers were tested in PS beams in 2009 \cite{PROF}.
In what follows, the test procedure of the 32$\times$48\,cm$^{2}$ ASUs prior to the m$^{2}$ prototype assembly is detailed.

\subsection{Detection threshold}

The study of the electronic characteristics consists in the determination of the noise level of the 1536 channels (64 channels/chip, 24 chips/ASU).
The number of hits recorded on a channel is measured as a function of the chip lowest threshold (common to the 64 chip channels).
The measured curve is called an S-curve: its inflexion point \textit{p} is the pedestal and its width \textit{w} relates to the noise level.
These quantities depend on the chip configuration like the shaping time or the preamplifier gain.
Once they are known, the working threshold is determined as \textit{p}$_{\rm k}$+5\textit{w}$_{\rm k}$ where the index ``k'' refers to the channel with the highest \textit{p}+5\textit{w} value.

\begin{figure}[h!]
\begin{centering}
\includegraphics[width=0.32\textwidth]{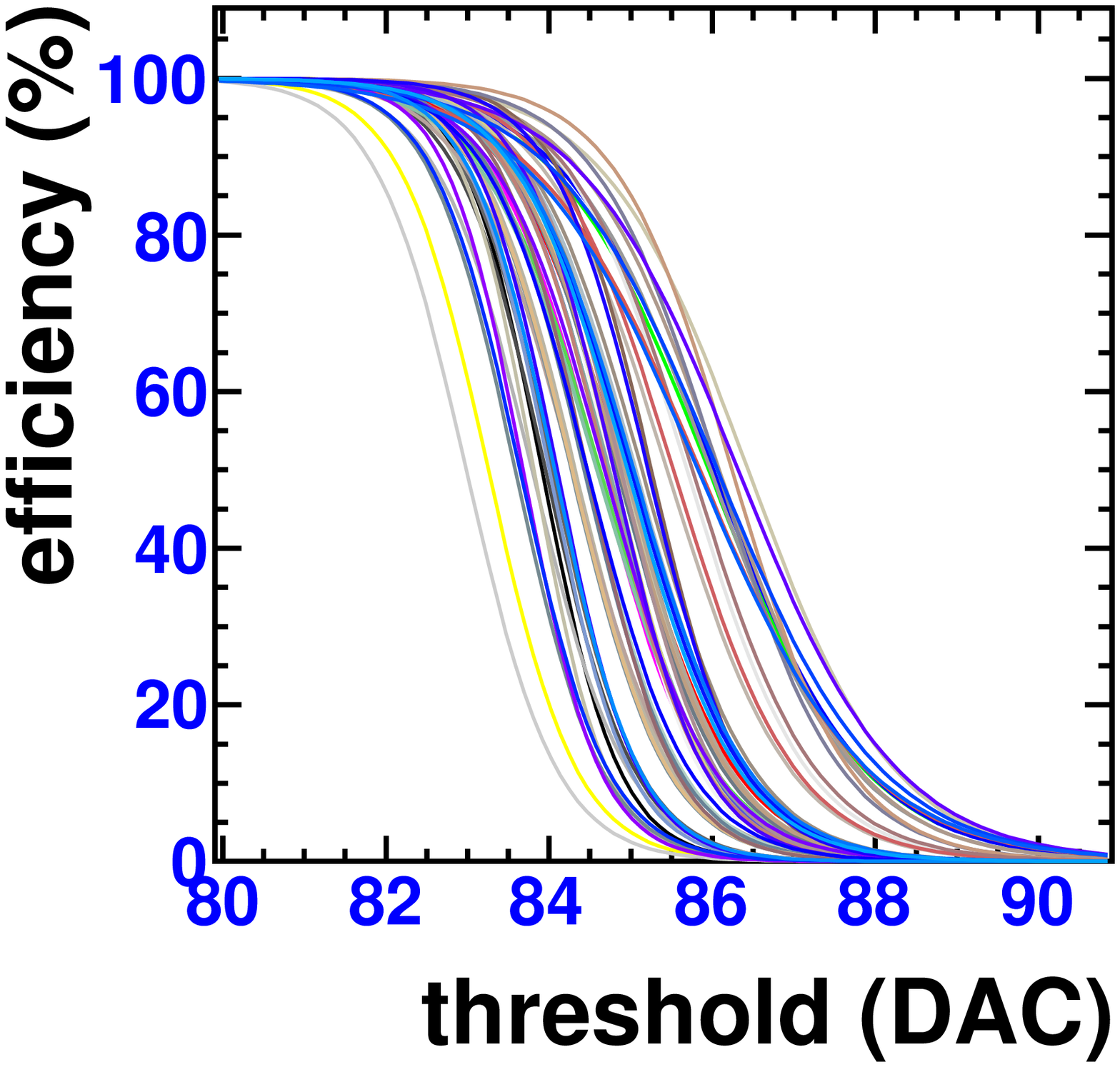}
\includegraphics[width=0.32\textwidth]{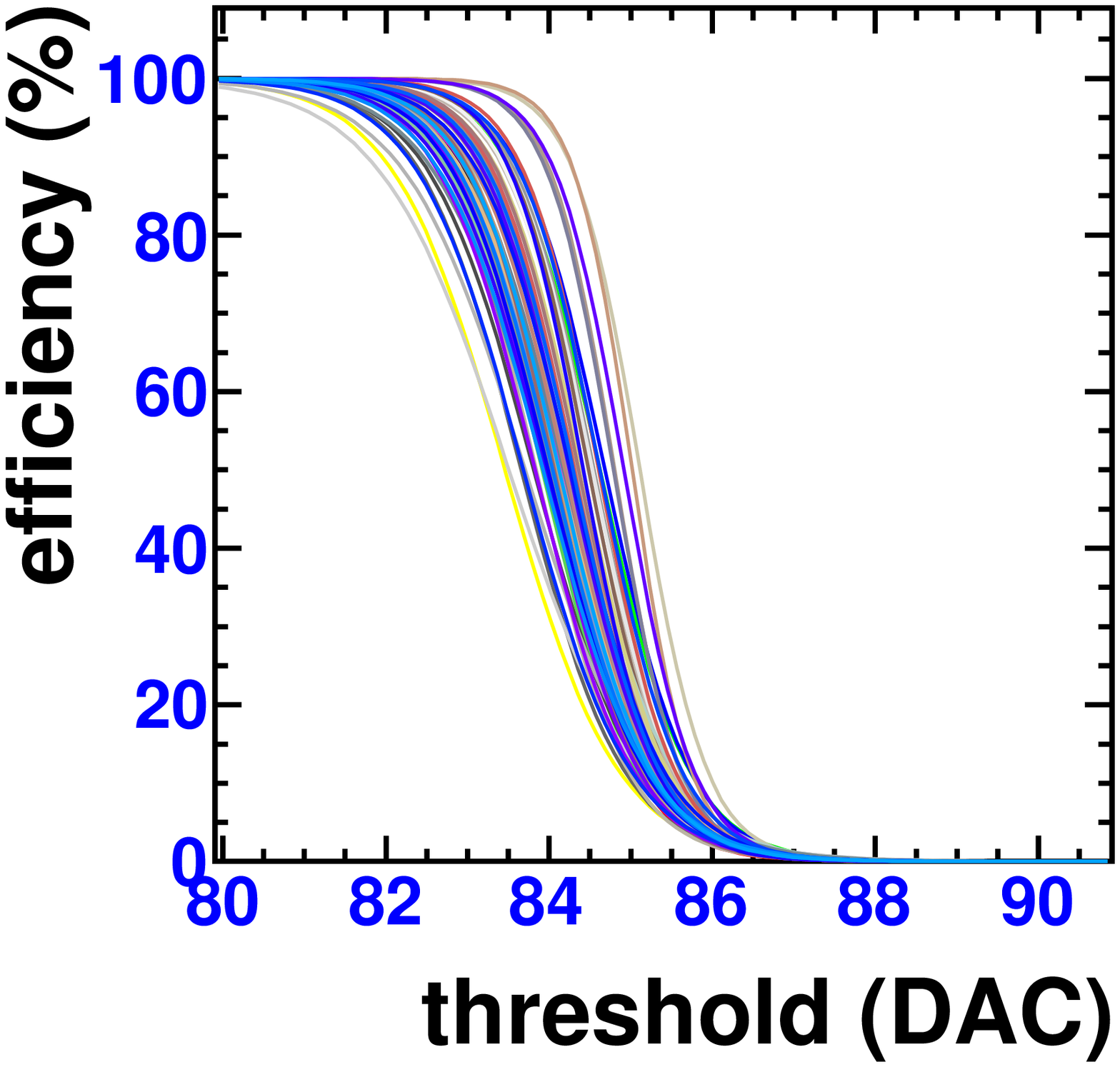}
\includegraphics[width=0.32\textwidth]{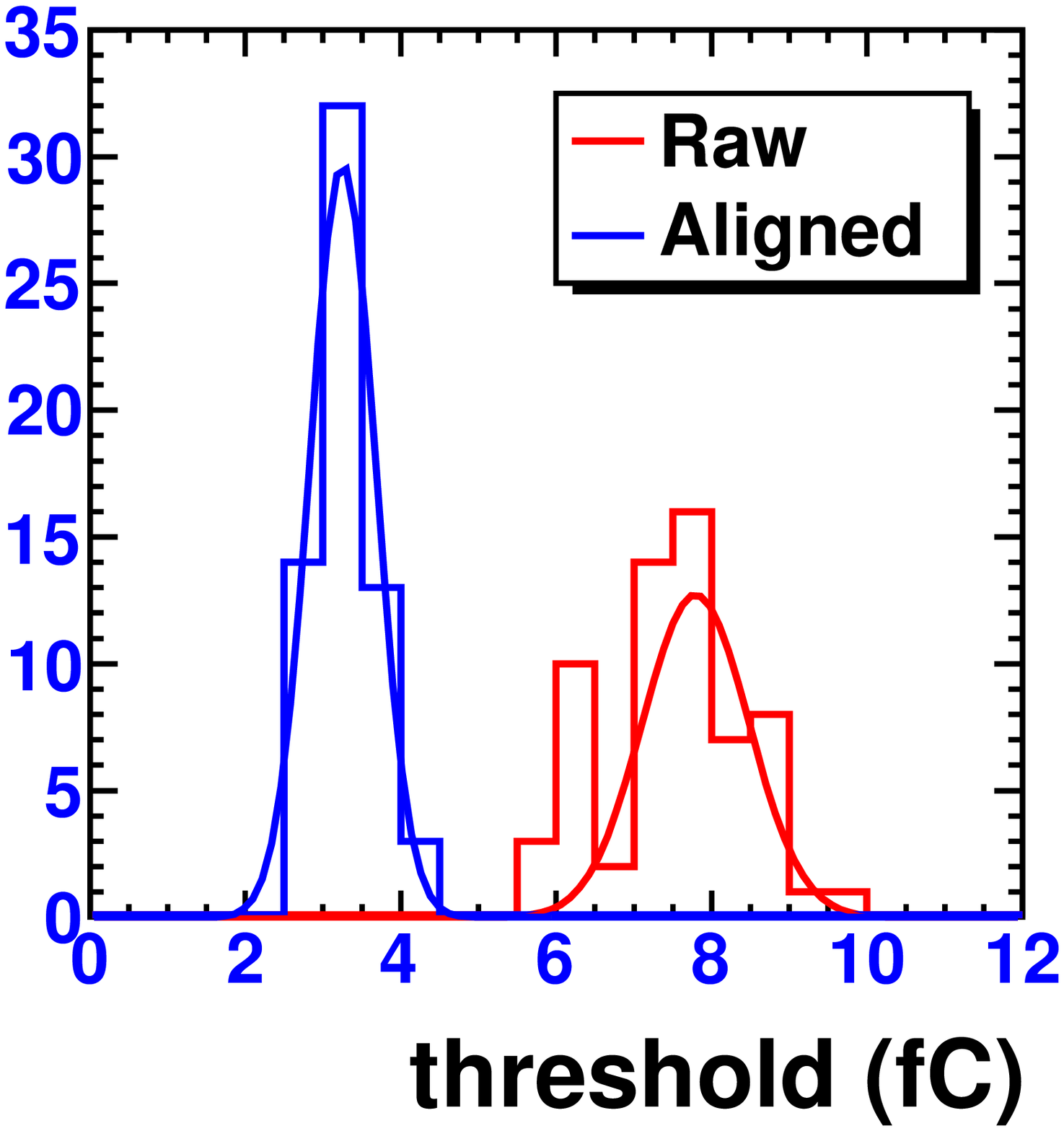}
\caption{S-curves measured on one chip without (left) and with gain corrections (center). Resulting charge threshold distributions (right).}
\label{scurves}
\end{centering}
\end{figure}

The dispersion of the S-curve parameters of the channels of a chip should be minimum in order to have the smallest detection threshold on each channel and also very little channel to channel threshold variations.
HARDROC2 chips feature an 8-bit adjustable gain preamplifier per channel.
Besides changing the preamplifier gain, it also changes the S-curve parameters and can eventually be used to align either the S-curve inflexion points \textit{p} or the S-curve endpoint \textit{p}+5\textit{w}.
The result of the S-curve alignment procedure is illustrated in Fig.\ref{scurves}.
When no gain corrections are applied, the difference between the minimum and maximum inflexion points is equal to 8.6 DAC units.
This corresponds to an average threshold of 7.6\,fC with an RMS of 1.0\,fC over the chip channels.
When the S-curve parameters are aligned (\textit{e.g.} \textit{p}+5\textit{w}), the S-curves fall over a DAC range of 3.4 units: the average threshold drops to 3.4\,fC with 0.4\,fC variations.

\subsection{Response uniformity}

The response uniformity of the ASU is determined mainly by the drift and amplification gap uniformity (primary ionisation yield and gas gain) and the preamplifier gain and charge threshold uniformity. When the most probable value of the preamplifier output signal is large compared to the threshold (\textit{e.g.} MPV/\textit{t}\,$\approx$\,22/1.5, cf. section \ref{gassiplex}), the threshold uniformity has very little influence and the preamplifier gains can be adjusted to correct for detector gap non-uniformity. As a first exercise, the detector gap effects were ignored and only the spread of the preamplifier gain distribution was minimized.

\begin{figure}[h!]
\begin{centering}
\includegraphics[width=0.35\textwidth]{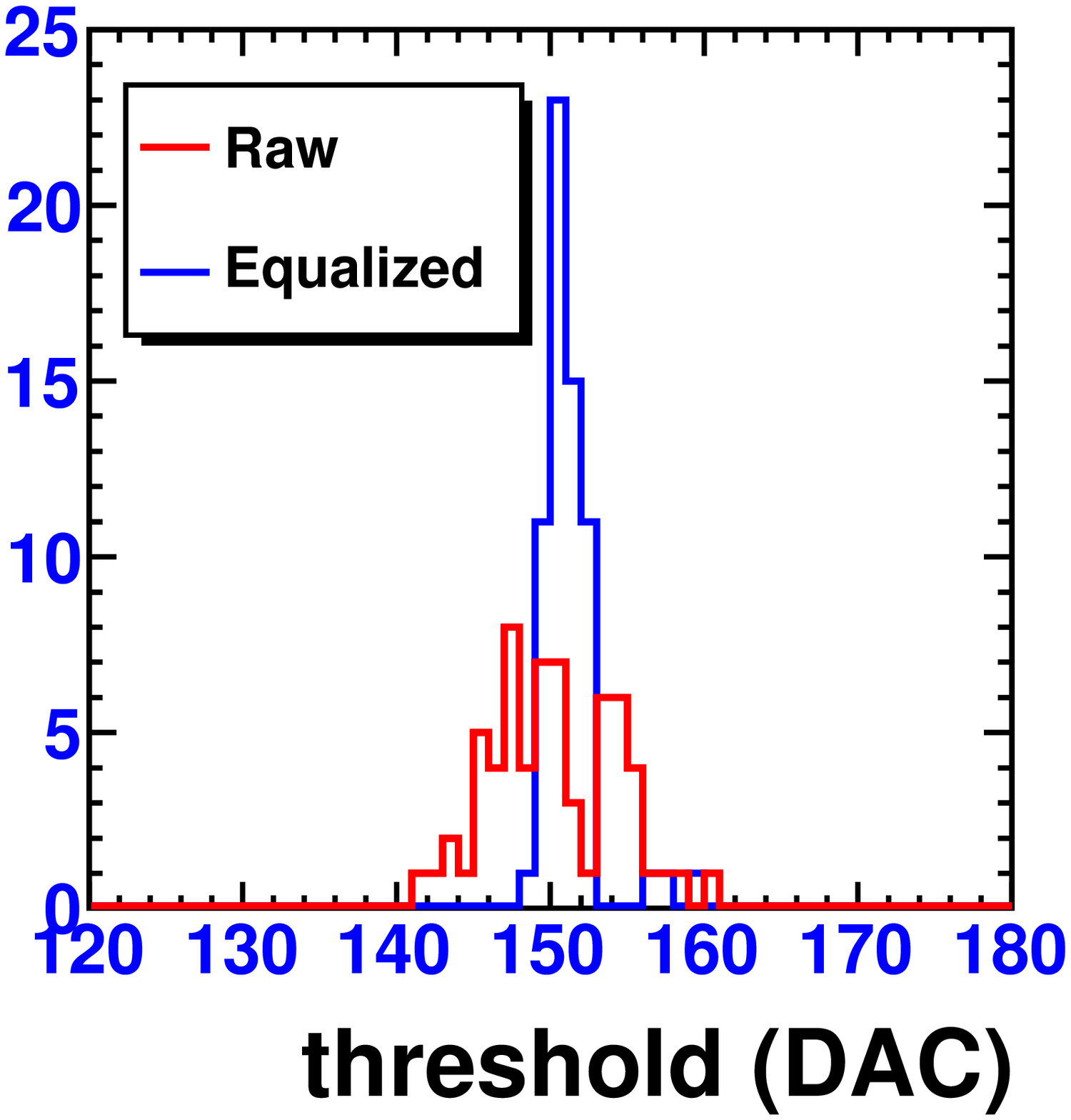}
\includegraphics[width=0.35\textwidth]{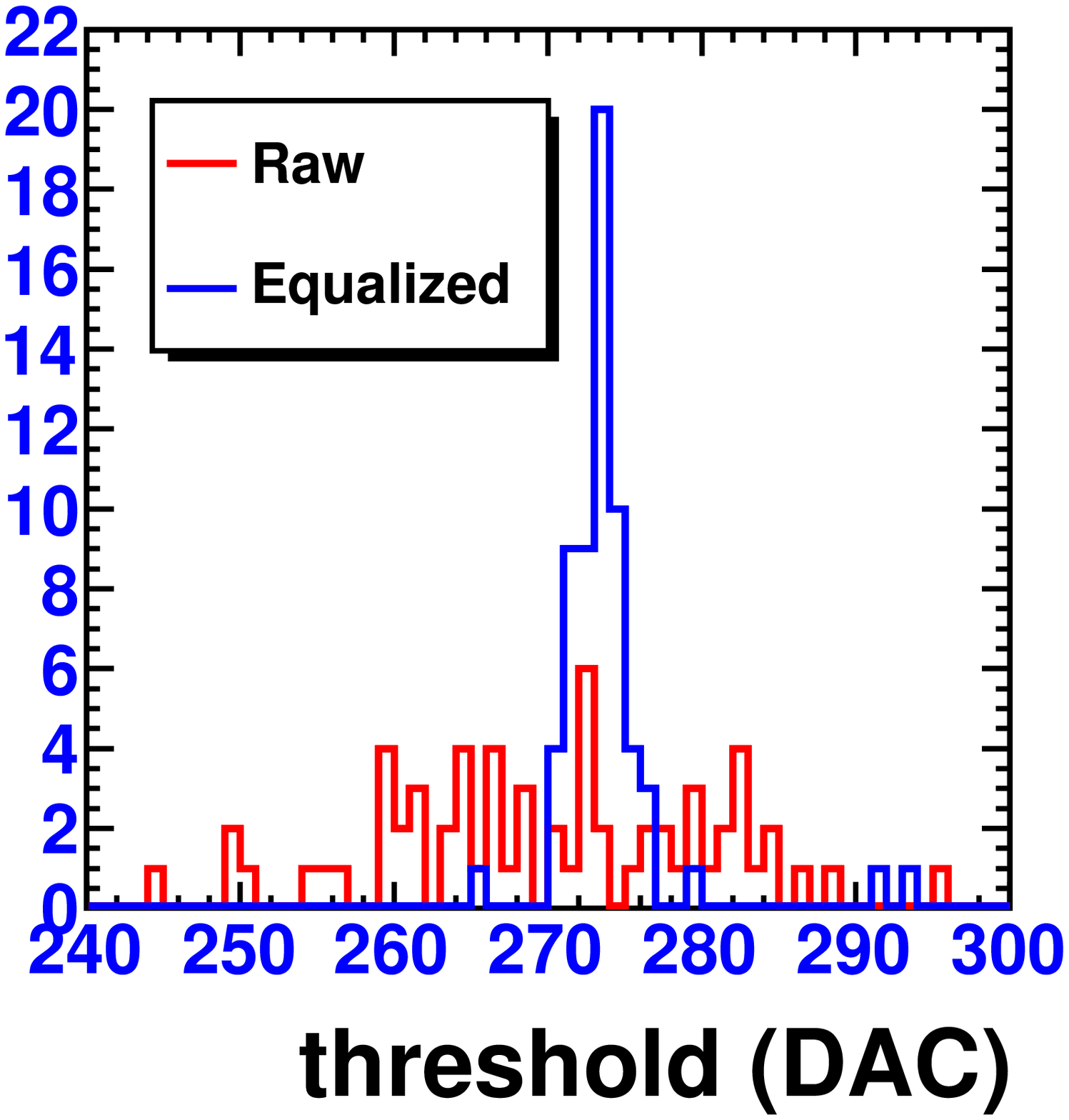}
\caption{Inflexion point distributions without and with gain equalization for 100\,fC (right) and 200\,fC (left) test charges.}
\label{equaliz}
\end{centering}
\end{figure}

A calibration input available on the HARDROC2 is used to inject a known charge to the preamplifier input of each channel.
The gain is calculated as the slope of the S-curves inflexion point versus injected charge trend \textit{p}(\textit{Q}$_{\rm in}$).
The gain of every channel is then corrected according to its measured raw value.
After this equalization procedure, the \textit{p}(\textit{Q}$_{\rm in}$) trends are parallel and the inflexion point distribution at a given charge exhibits a reduced dispersion. This can be seen in Fig.\ref{equaliz} (left) where the distribution \textit{p}(100\,fC) with and without equalization is shown. The natural dispersion of 4\,\% is reduced to 1.2\,\%. This procedure also works at different input charges as is illustrated in Fig.\ref{equaliz} (right) at 200\,fC.

\subsection{Tests with X-rays}

A dedicated gas chamber has been fabricated to perform tests of the ASUs with $^{55}$Fe 5.9 keV X-rays.
The chamber can house one ASU. It has a perforated aluminium cover onto which is glued a cathode foil with a few nm thin conductive deposit on one side.
It is therefore transparent to $^{55}$Fe quanta which are almost all absorbed in the 3\,cm thick drift gap.

When absorption by the photo-electric effect on an argon atom, the quanta creates a point-like cloud of 230 electrons which drift towards the mesh.
During electron multiplication, the signal induced at the preamplifier input is amplified. It is then shaped and fed to a discriminator.
If higher than the threshold, the signal is detected and a hit is recorded.
Because the digital clock frequency is much higher than the quanta conversion rate in the chamber gas, there is basically no dead time and the number of hits \textit{N}$_{\rm hit}$ recorded in a given time is an indication of the detection efficiency.

\begin{figure}[h!]
\begin{centering}
\includegraphics[width=0.49\textwidth]{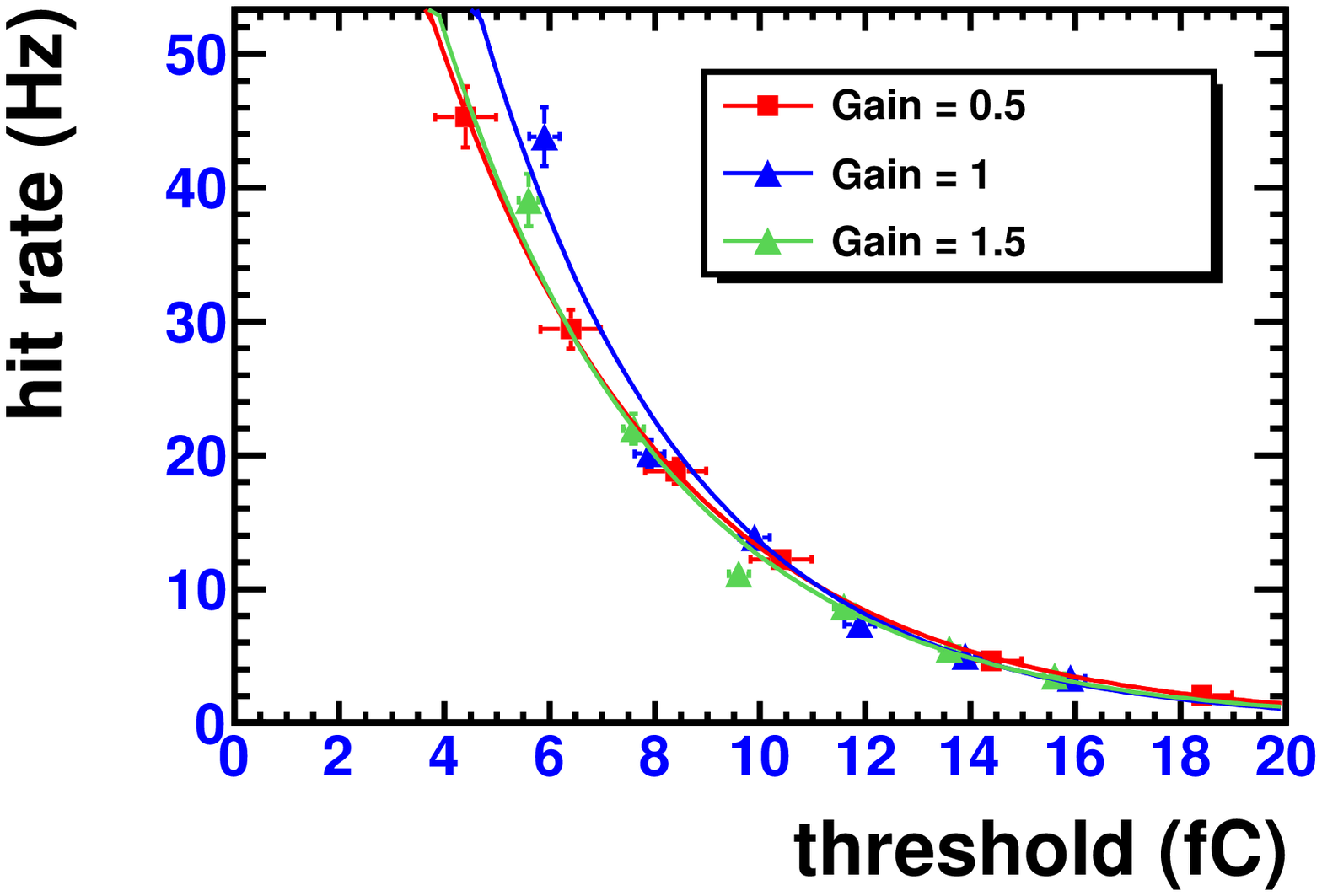}
\includegraphics[width=0.49\textwidth]{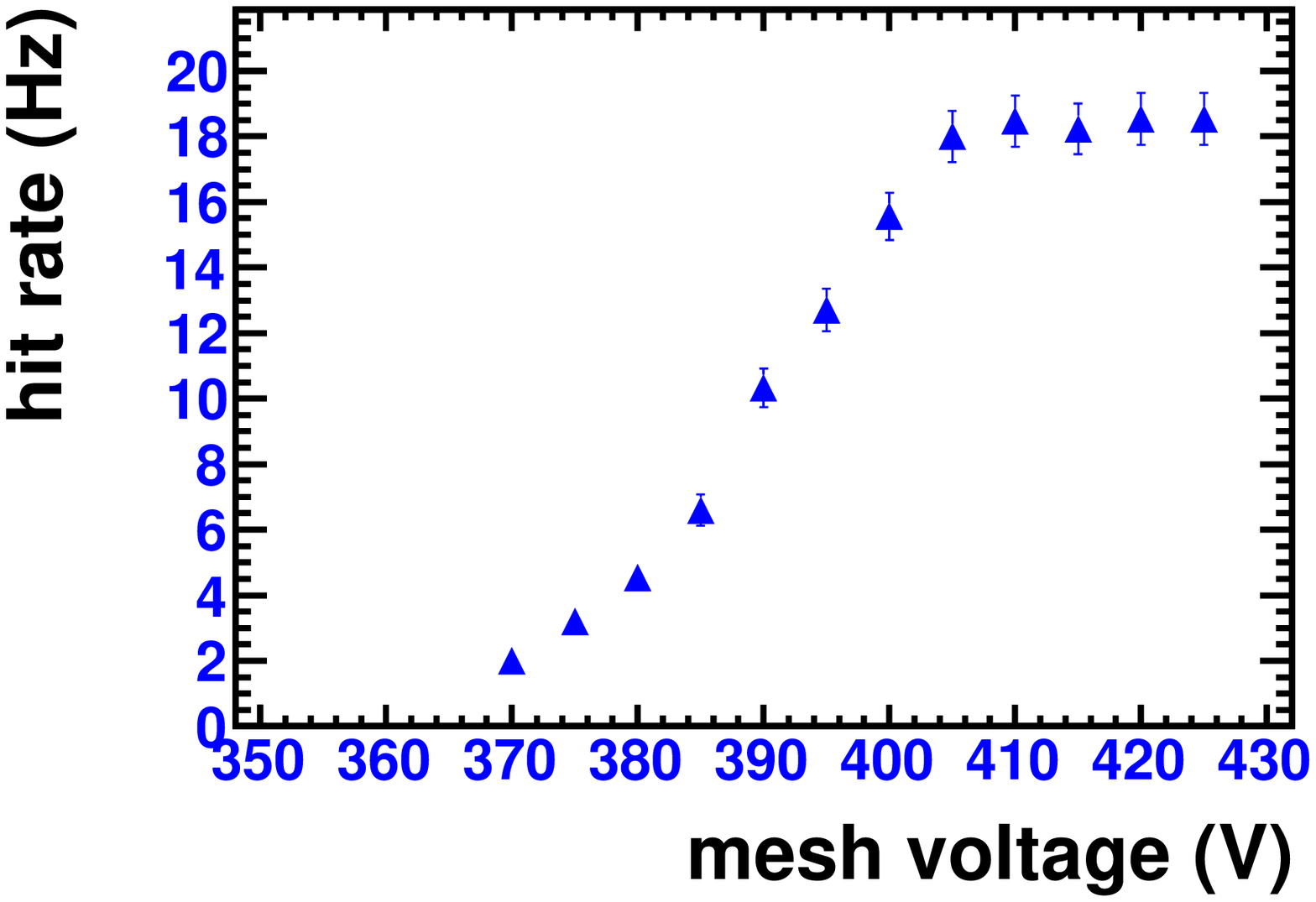}
\caption{Number of hits per unit time from $^{55}$Fe quantum conversions as a function of chip threshold at different preamplifier gains (left) and mesh voltage (right).}
\label{testbox}
\end{centering}
\end{figure}

Such a counting experiment was performed on a given pad for different chip configurations and bias voltages, the counting time was 30\,s.
The effect of the threshold can be appreciated in Fig.\ref{testbox} (left) where \textit{N}$_{\rm hit}$ falls off at increasing threshold.
This trend does not depend on the preamplifier gain (measurements at gains of 0.5, 1 and 1.5) because the signal to noise ratio is constant.\\
The multiplication factor is an exponential function of the mesh voltage which should strongly influence \textit{N}$_{\rm hit}$.
This is illustrated in Fig.\ref{testbox} (right) where the first $^{55}$Fe quanta are being counted at 370\,V. A detection plateau is reached at 410\,V at a gas gain of 10$^{4}$.

\section{Simulation studies}

\subsection{Introduction}

At an ILC detector with PFA-oriented calorimeters, the measurement of the energy of jets relies both on the tracking and the calorimetry.
In particular, the capability to group calorimeter hits together and to separate individual showers within jets will be crucial \cite{TOMP}.
Such a performance can be investigated with simulation programs.
Also, the influence of the absorber material, particle energy, calorimeter depth and segmentation on the energy resolution, linearity, shower profiles and leakage can be studied with the simulation. The comparison between analog and digital readout is reported below.
For further studies, the reader is referred to \cite{JAN}.

\subsection{Analog and digital readout}

The geometry of the hadronic calorimeter is based on the one proposed for the SiD detector \cite{SID} but twice deeper (9 $\lambda$ instead of 4.5 $\lambda$). The lateral size is 1$\times$1\,m$^{2}$ while the depth varies from 1.70 to 2.39\,m depending on the absorber material. Monte Carlo data for negative pions were generated by a GEANT4-based simulator SLIC with LHEP physics list. When calculating the performance with a digital readout, a threshold of 10\,\% of the MIP MPV is applied to the energy deposited in each 1\,cm$^{2}$ cells.

\begin{figure}[h!]
\begin{centering}
\includegraphics[width=0.49\textwidth]{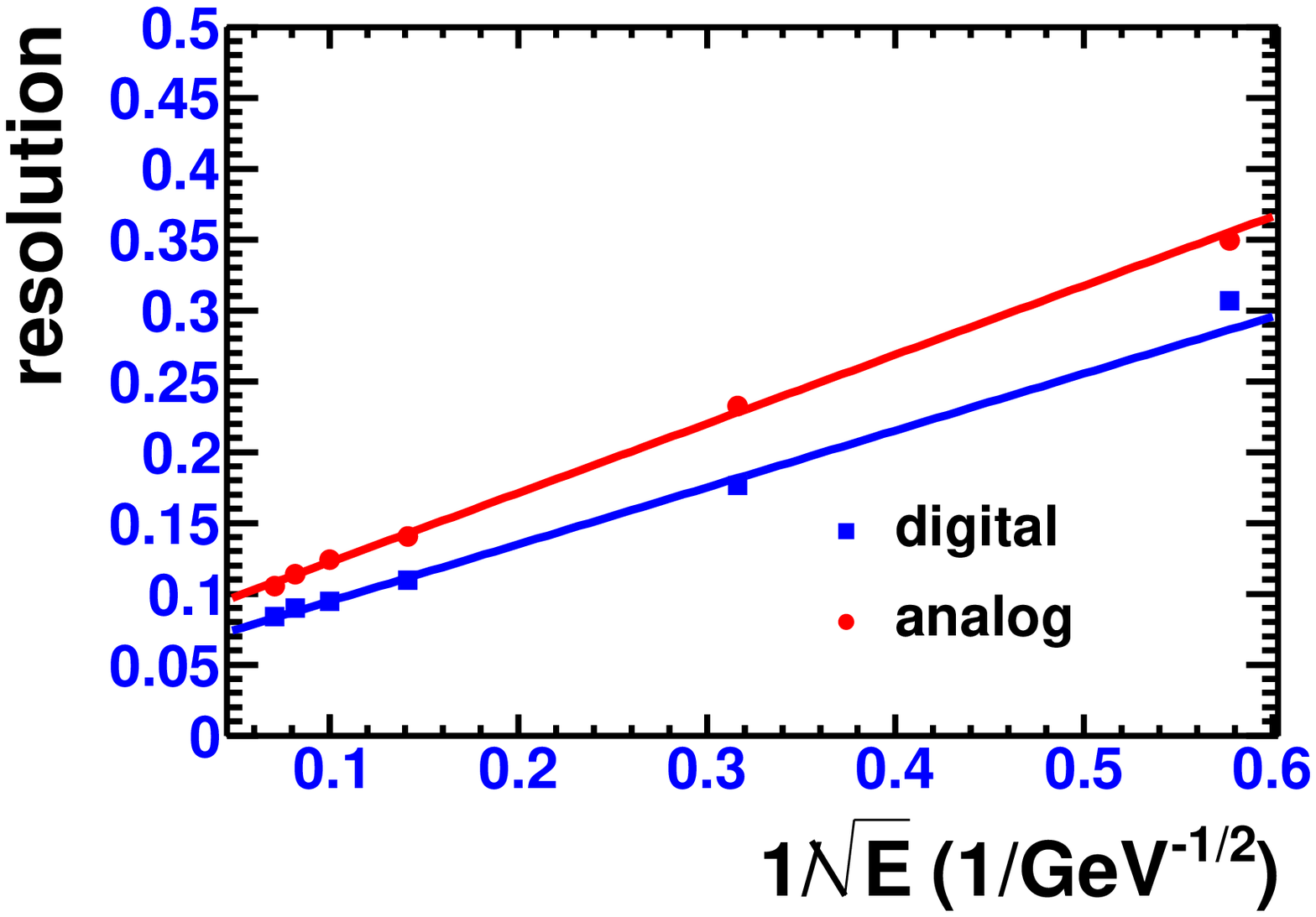}
\includegraphics[width=0.49\textwidth]{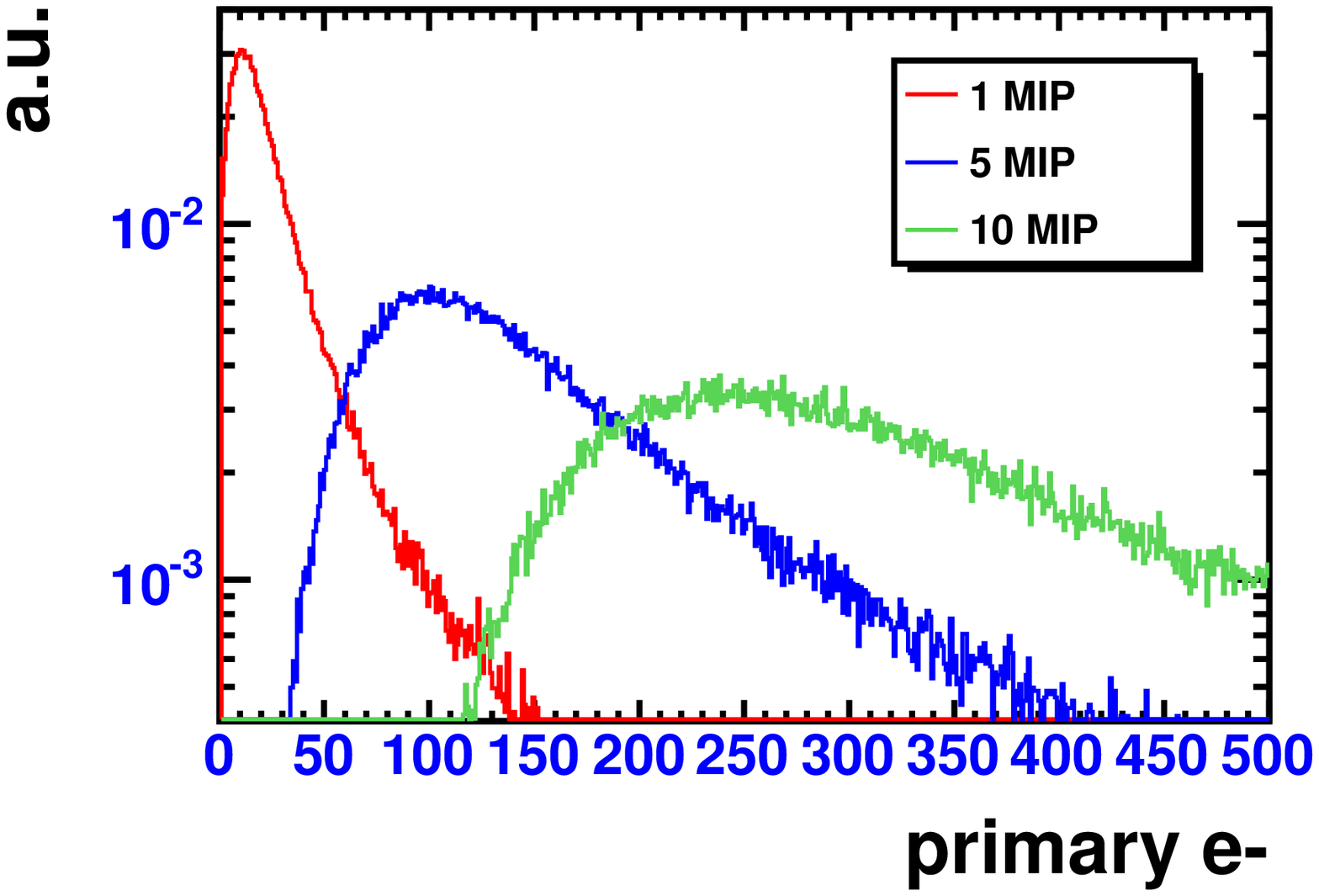}
\caption{Energy resolution for pions with a digital (0.1 MIP threshold) and analog readout (left). Primary charge distributions for various number of traversing MIPs (right). The distributions are calculated with a parametrization of a measured Landau distribution \cite{AMB} and a primary electron density taken from \cite{SAULI}.}
\label{reso}
\end{centering}
\end{figure}

\vspace{0.5cm}
The energy resolution as a function of pion energy of a stainless steel calorimeter and different readouts is displayed in Fig.\ref{reso} (left).
At low pion energy, the energy resolution is dominated by Landau fluctuations which are suppressed when applying a threshold.
As a result, the digital readout provides a better resolution.
At higher energy, however, the number of hits saturates due to the finite size of the anode pads and a loss of linearity is observed in the digital case.
Yet, the digital readout shows a superior resolution.
The loss of linearity can in principle be overcame with a semi-digital readout where more than one threshold would be applied.
In contrast with RPCs, Micromegas is well suited for a semi-DHCAL because no saturation effects take place in the gas volume and the read out signals are proportional to the primary ionisation yield (Fig.\ref{reso} (right)). The determination of the charge thresholds and their corresponding weights is the subject of current investigation.

\section{Conclusion}

The Micromegas semi-DHCAL 1\,m$^{3}$ project has been surveyed.
At detection thresholds of a few fC, high efficiency and low hit multiplicity were measured with small chambers and analog readout.
The challenge is to maintain these performance on larger area detectors with PCB embedded digital electronics (Active Sensor Units).
The technique to manufacture such compact detectors is well controlled and can be used for the production of m$^{2}$ planes.
In parallel to that, a strong effort is made on the simulation. This aspect of the project is essential for designing and predicting the main performance of a Micromegas semi-DHCAL.

The 1\,m$^{2}$ Micromegas prototype will be composed of 6 ASUs which are individually tested in the laboratory prior to the assembly.
A careful study of the front-end chip noise characteristics has been carried out to minimize the detection threshold and a gain equalization procedure has been established in order to improve the response uniformity. Finally, the overall functioning of the ASUs has been verified by performing X-ray tests in a dedicated gas chamber. At the moment of writing, the first 1\,m$^{2}$ Micromegas prototype has been assembled and tested in a muon beam at the CERN/SPS facility in June/July 2010. Preliminary results are promising and the data analysis is on-going.

\section{Acknowledgments}

The author would like to thank Rui de Oliveira and Olivier Pizzirusso from the CERN EN-ICE-DEM group for the lamination of the Bulk Micromegas mesh on the boards as well as D. Atti\'e, P. Colas and W. Wang from Irfu/Saclay for their participation to the 2009 test beams. Finally, the fruitful collaboration between the LAL/Omega and LAPP/LC group should be underlined: many thanks to N. Seguin and C. de la Taille for their participation on the test of the first HARDROC2 equipped ASUs and on the recent ASIC developments.

\end{document}